\documentclass[prb,twocolumn,showpacs,floats,eqsecnum,amsmath,amssymb]{revtex4}
\usepackage{epsf}
\usepackage{times}
\usepackage[dvips]{graphicx}
\begin{document}

\title{Magnetic interference pattern in planar SNS Josephson
       junctions}

\author{G. Mohammadkhani$^1$, M. Zareyan$^1$, and Ya. M. Blanter$^2$}

 \affiliation{$^1$Institute for Advanced Studies in Basic Sciences (IASBS),
 P. O. Box 45195, Zanjan 45195, Iran\\
 $^2$  Kavli Institute of NanoScience,
        Delft University of Technology, Lorentzweg 1, 2628 CJ Delft, The
        Netherlands}

\date{\today}
\begin{abstract}
We study the Josephson current through a ballistic normal metal
layer of thickness $D$ on which two superconducting electrodes are
deposited within a distance $L$ of each other. In the presence of
an ({\it in-layer}) magnetic field we find that the oscillations
of the critical current $I_c(\Phi)$ with the magnetic flux $\Phi$
are significantly different from an ordinary magnetic interference
pattern. Depending on the ratio $L/D$ and temperature,
$I_c(\Phi)$-oscillations can have a period smaller than flux
quantum $\Phi_0$, nonzero minima and damping rate much smaller
than $1/\Phi$. Similar anomalous magnetic interference pattern was
recently observed experimentally.
\end{abstract}

 \maketitle

\section{Introduction}
Existence of a supercurrent in a Josephson junction is the manifestation of the
interference between the macroscopic wave functions (superconducting order
parameters) of the two contacted superconductors. The quantum interference can be
modulated by an external magnetic field applied to the junction. As the
result the critical(maximum) supercurrent $I_c(\Phi)$ shows a well-known Fraunhofer
diffraction pattern-like dependence on the magnetic flux $\Phi$ penetrating the
junction area. In a superconductor-insulator-superconductor (SIS)
junction the critical current,
\begin{equation} \label{Fraunhofer}
I_c (\Phi) =I_c(0) \frac{\sin(\pi \Phi/\Phi_0)} {\pi \Phi/ \Phi_0},
\end{equation}
oscillates with the period of flux quantum
$\Phi_0 = \pi \hbar c/e$ and an amplitude decreasing as $1/\Phi$
~\cite{JosephsonRMP1964,RowellPRL1963}. The main features of the
effect, {\em i.e.}
damped oscillations of $I_c$ with the magnetic flux, take place in other types of
Josephson weak links, however the detailed behavior, including the period of the
oscillations and the rate of damping, depends on the geometry as well as the nature
of the weak link.

In a wide SNS (N being a normal metal layer) junction $I_c$ has a similar magnetic
interference pattern as SIS systems\cite{Antsygina1975}. On the other hand, Heida
{\it et al.,}~\cite{HeidaPRBR1998} investigating S-two-dimensional-electron-gas-S
(S2DEGS) junctions of comparable width $W$ and length $L$, have measured a 2$\Phi_0$
periodicity of the critical current instead of the standard $\Phi_0$ periodicity. The
first explanation of this finding was due to Barzykin and Zagoskin
\cite{Zagoskin1997}, who considered a S2DEGS junction with perfect Andreev
reflections at NS interfaces and both absorbing and reflecting lateral boundaries,
obtained a 2$\Phi_0$ periodicity in the limit ${L/W\rightarrow\infty}$ ({\em i.e.} in
the limit of the point-contact geometry). Later, Ledermann {\it et al.,}
\cite{LedermannPRBR1999} considered more realistic reflecting boundaries at the edges
and found that in the limit of strip geometry ($L/W\simeq1$) the periodicity of the
critical current changes from $\Phi_0$ to 2$\Phi_0$ as the flux through the junction
increases. In general, increase of the periodicity was attributed to the nonlocality
of the supercurrent density in hybrid NS structures.

In this paper we report on a new type of magnetic interference pattern  in a planar
SNS junction. The SNS junction studied below consists of a thin normal metal layer of
thickness $D$ on which two superconducting electrodes are deposited in a distance $L$
of each other (Fig.~\ref{mzbfig1}). An external magnetic field is
applied in plane of N-layer
perpendicular to the direction of the Josephson current flow. Such a Josephson setup
was recently used in the experiment by Keizer~{\it et al.}\cite{KlapwijkNature2006} to
investigate the Josephson supercurrent through a S-half-metallic-ferromagnet-S
(SHMFS) junction. They used NbTiN superconducting electrods on top of a thin layer of
CrO$_2$, which is a fully polarized (half-metallic) ferromagnet, {\em
i.e.} it supports only one spin direction of electrons. Surprisingly,
Josephson supercurrent was detected for the
junction length $L \sim {\rm300 nm}$ in spite of strong pair breaking in CrO$_2$ which
is expected to suppress all singlet superconducting correlations.
Ref. \onlinecite{KlapwijkNature2006} also reports measurement of the magnetic
interference pattern with magnetic field applied in plane of HMF-layer.  It is
found that $I_c(\Phi)$ oscillates with the in-plane flux $\Phi$ with a period of
order $\Phi_0$. In contrast to a standard magnetic interference
pattern (\ref{Fraunhofer}), $I_c$ has
nonzero values at the minima and the amplitude of the oscillations decreases rather
slowly compared to $1/\Phi$.

The long range superconducting proximity in HMF can be explained in terms of the
triplet superconducting correlations generated at spin-active HMFS-interfaces as
the result of interplay between the singlet superconducting correlations and a
noncollinear magnetization inhomogeneity~\cite{BergeretPRL2001,BergeretRMP2005}.
While the singlet superconducting correlations are destroyed over a short distance of
the Fermi wave-length, the triplet components can survive over
distance of the order of
the normal coherent length $\xi_N=v_{\rm F}/2\pi T$, with $v_{\rm F}$ being the Fermi
velocity. In Ref. \onlinecite{KlapwijkNature2006} by investigating the critical
supercurrent for different distances between the electrodes it was concluded that the
length dependence is the same as in nonmagnetic SNS junctions. This
strongly suggests indeed that
triplet correlations are responsible for the observed Josepshson
current.

The aim of the present paper is to investigate $I_c(\Phi)$ in such a
planar SNS junction. Whereas this problem is interesting by itself, we
also believe that it is relevant for the understanding of the
experimental results of Ref. \onlinecite{KlapwijkNature2006}. Indeed,
penetration of triplet correlations in SHMFS junction is similar to an
ordinary
(singlet) superconducting proximity in SNS systems, {\em i.e.} both
decay exponentially
within the length scale $\xi_N$. By noting this fact we will use the quasiclassical
Green's function formalism to investigate the magnetic flux dependence of the
supercurrent in the corresponding planar SNS junction (see Fig. \ref{mzbfig1}). We
find that the magnetic interference pattern is significantly different from that of
the standard one. The period of the oscillations can be smaller than $\Phi_0$
depending on the length-to-thickness ratio $L/D$. The period tends to $\Phi_0$ at
higher magnetic fluxes and also for very large $L/D$. We also obtain the two
anomalous features observed in the experiment: the amplitude of the oscillations has
a rather slow decrease with $\Phi$ compared to the standard SIS case
(\ref{Fraunhofer}), and the critical current as the function of the
flux, $I_c(\Phi)$,
at low temperatures can have finite minima when the total flux $\Phi$ is integer or
non-integer multiplies of the $\Phi_0$ depending on the period of the
oscillations.

In Section~\ref{sec2} we introduce our model of a ballistic
planar SNS contact and present solutions of the Eilenberger
equation for the quasiclassical Green's functions of a given
electronic trajectory. Introducing the effect of the in-plane
magnetic field through a gauge invariant phase we obtain the
expression of the critical supercurrent as a function of the
magnetic flux.  Section \ref{sec3} is devoted to the analysis of
the $I_c(\Phi)$ in terms of $L/D$ for different temperatures.
In Section~\ref{sec4} we present the conclusion.

\section{Josephson current in SNS junction with an in plane magnetic field}
\label{sec2}

\begin{figure}
\centerline {\hbox{\epsfxsize=2.7in \epsffile {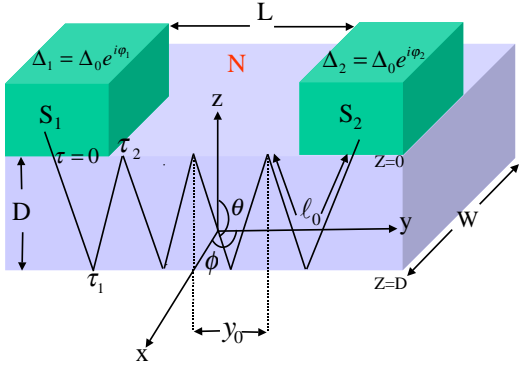}}}
\caption{(Color online), Schematic of the SNS junction. The
current flow between two superconductors (${\rm S_{1,2}}$) through
a Normal layer (N). The NS interfaces are perfectly transparent.}
\label{mzbfig1}
\end{figure}

In this section, we calculate the Josephson current for a clean SNS junction in the
presence of an external magnetic field $\mathbf{H}$. The setup is schematically shown
in Fig. \ref{mzbfig1}. It consists of a normal metallic (N) layer of the thickness
$D$ and width $W$ on which two superconducting electrodes are deposited in a distance
of $L$. This planar Josephson structure was studied experimentally in Ref.
\onlinecite{KlapwijkNature2006} with a half metallic N layer. A phase difference
$\varphi$ between order parameters of the superconductors drives a Josephson
supercurrent through the parts of N layer underneath the superconductors and the
junction N part ($-L/2\leq y\leq L/2$). The magnetic field
$\mathbf{H}=-H\hat{\textrm{x}}$ is applied in the plane of N layer perpendicular to
the direction of supercurrent flow. We consider a clean structure with all dimensions
$L, W, D$ being smaller than the electronic impurity mean free path $\ell_{\rm imp}$
and ideally transparent NS-interfaces. At the same time, the Fermi wavelength is small
compared to $L, W, D,$ and the superconducting coherence length $\xi_0=v_{\rm
F}/2\Delta_0(T=0)$ (we use the system of units with
$\hbar=k_{\text{B}}=1$).
Under these conditions the electronic properties of the system can be
derived from the Eilenberger
equations for the semiclassical matrix Green's function $\hat{\textit{g}}$,
\begin{equation}
-v_{\rm F}
\mathbf{n}.\nabla\hat{\textit{g}}=\omega_{n}[\hat{\tau}_{3},\hat{\textit{g}}~]+[\hat
{\Delta}(\textbf{r}),\hat{\textit{g}}~],
 \label{JC:eq1}
\end{equation}
The matrix Green's function
\begin{eqnarray*}
\hat{\textit{g}}=\left(\begin{array}{cc}
\textit{g}_{\omega_{n}}& f_{\omega_{n}}\\
f^{\dagger}_{\omega_{n}}& -\textit{g}_{\omega_{n}}
\end{array}\right),
\end{eqnarray*}
where the normal $\textit{g}$ and anomalous $f$ Green's functions depend on the
Matsubara frequency $\omega_{n}=\pi T(2n+1)$, on the coordinate $\textbf{r}$, and the
direction of motion $\textbf{n}$; $\hat{\tau}_{i}$ (i=1,2,3) denotes the Pauli
matrices in the Nambu space, and the matrix
\begin{eqnarray*}
{\hat{\Delta}(\textbf{r})}= \left(\begin{array}{cc}
\begin{array}{c}
0
\end{array}
&
\begin{array}{c}
\Delta(\textbf{r})
\end{array}
\\
\begin{array}{c}
\Delta^{\ast}(\textbf{r})
\end{array}
&
\begin{array}{c}
0
\end{array}
\end{array}\right)
\end{eqnarray*}
represents the superconducting order parameter $\Delta(\textbf{r})$. The matrix
Green's function $\hat{\textit{g}}_{\omega_{n}}$ satisfies the normalization
condition,
\begin{align}
\hat{\textit{g}}^2=1&,&\textit{g}_{\omega_{n}}^{2}+f_{\omega_{n}}f^
{\dagger}_{\omega_{n}}=1,
 \label{JC:eq2}
\end{align}
where $f^{\dagger}_{\omega_{n}}(\textbf{r},
\mathbf{n})=f^{\ast}_{\omega_{n}}(\textbf{r},-\mathbf{n})$. In components, the
Eilenberger equations have the form
\begin{eqnarray}
-v_{\rm F}\mathbf{n}.\nabla\textit{g}_{\omega_{n}}&=&\Delta(\textbf{r})f^{\dagger}_
{\omega_{n}}-\Delta^{\ast}(\textbf{r})f_{\omega_{n}},
\label{JC:eq3}\\
-v_{\rm F}\mathbf{n}.\nabla
f_{\omega_{n}}&=&2\omega_{n}f_{\omega_{n}}-2\Delta(\textbf{r})
\textit{g}_{\omega_{n}}.
 \label{JC:eq4}
\end{eqnarray}
We solve these equations along an electronic quasiclassical trajectory (shown in Fig.
\ref{mzbfig1}) which is parameterized by $-\infty \leq \tau \leq
\infty$~\cite{ZareyanPRL2001,ZareyanPRB2002}. Assuming a weak external magnetic field
we neglect its effect on the orbital motion of the quasiparticles. The magnetic field
then will have only phase effect which we will include by introducing the gauge
invariant phase as
\begin{eqnarray}
\varphi=\varphi_0-\frac{2\pi}{\Phi_0}\int_{0}^{\ell}\mathbf{A}.\mathbf{n}d\tau,
\label{JC:eq5}
\end{eqnarray}
where $\varphi_0$ is the phase difference between two superconductors in the
absence of the magnetic field, the second part is the phase accumulated by the
quasiparticle on the trajectory due to the magnetic field
$\mathbf{H}=\nabla\times\mathbf{A}$, with $\ell$ being the length of
the trajectory inside N-layer. The vector potential is taken
to be $\mathbf{A}=-Hz\hat{\textrm{y}}$.

A typical trajectory consists of three parts: the part extended from bulk of ${\rm
S_1}$ ($\tau=-\infty$) to a point at ${\rm NS_1}$-interface $\tau=0$, the part inside
N layer ($0<\tau<\ell$) which extends from a point at ${\rm NS_1}$-interface to a
point at ${\rm NS_2}$-interface and the last part which extends from the point
$\tau=\ell$ to the bulk of ${\rm S_2}$ ($\tau=\infty)$. Our equations
are supplemented by the boundary conditions which determine the values of
Green's functions in the bulk ${\rm S_1}$ and ${\rm S_2}$,
\begin{eqnarray}
f_{\omega_{n}}(\tau=\mp\infty)&=&\frac{\Delta_0(T)
  \exp(i\varphi_{1,2})}{\Omega_n},
\label{JC:eq6}\\
\textit{g}_{\omega_{n}}(\tau=\mp\infty)&=&\frac{\omega_n}{\Omega_n} \ ,
 \label{JC:eq7}
\end{eqnarray}
where $\Omega_{n}=\sqrt{\Delta_{0}^{2}+\omega_{n}^{2}}$, and $\Delta_{0}(T)$ is the
temperature-dependent superconducting gap. We neglect the variation of the order
parameter close to NS-interfaces inside the superconductors and approximate the order
parameter by the step-function, $\Delta(\textbf{r})=\Delta_0
e^{i\varphi_{1}}\theta(\tau)+\Delta_0e^{i\varphi_{2}}\theta(\tau-L)$. We can obtain
the Green's function $\textit{g}_{\omega_{n}}$ on a trajectory inside N that is
constant. It depends only on the length of that trajectory $\ell$ and the phase
difference $\varphi_0$, which is given by
\begin{eqnarray}
\textit{g}_{\omega_{n}}= \tanh[\omega_{n}\ell/v_{\rm
F}+i\varphi_0/2+\mbox{\rm arcsinh}\ (\omega_{n}/\Delta_{0})] \ .
 \label{JC:eq8}
\end{eqnarray}
Note that in the presence of the magnetic field, the phase difference $\varphi_0$ is
replaced by $\varphi$ (see Eq. (\ref{JC:eq5})).

The supercurrent density can then be obtained by averaging (\ref{JC:eq8}) over all
different possible classical trajectories. This corresponds to an averaging over
Fermi velocity directions. In the presence of the planar magnetic field, we find
\begin{eqnarray}
\mathbf{j(\mathbf{r})}=e\mathcal{N}(0)Tv_{\rm F}\sum_{\omega_{n}} \int \mathbf{n}
\mbox{\rm Im}\ \textit{g}_{\omega_{n}}(\ell,v_{\rm
  F})\sin\theta d\theta d\phi \ .
 \label{JC:eq9}
\end{eqnarray}
Here $\mathcal{N}(0)$ is the density of states at the Fermi surface and
$\mbox{Im}\ \textit{g}_{\omega_{n}}(\ell,v_{\rm F})$ denotes the
imaginary part of the normal component of the matrix Green's function,
given by
\begin{eqnarray}
\nonumber \mbox{Im}\ \textit{g}_{\omega_{n}}&=&\frac{\Delta_{0}^{2}(T)\sin\varphi}
{(\Omega_{n}^2+\omega_{n}^2)\cosh\chi+2\Omega_{n}\omega_{n}
\sinh\chi+\Delta_{0}^{2}\cos\varphi},\\
 \label{JC:eq10}\\
\chi&=&\frac{\omega_{n}\ell}{\pi T_c\xi_0}.
 \label{JC:eq11}
\end{eqnarray}

To calculate the integral of the vector potential along the quasiclassical
trajectories, we split it into the segments as shown in Fig. \ref{mzbfig1},
\begin{eqnarray}
\nonumber &&\int_{0}^{\ell}\mathbf{A}.\mathbf{n}d\tau=\int_{0}^{\tau_1}\mathbf{A}.
\mathbf{n}d\tau+
\int_{\tau_1}^{\tau_2}\mathbf{A}.\mathbf{n}d\tau+...\\
&&+\int_{\tau_{n-2}}^{\tau_{n-1}}\mathbf{A}.\mathbf{n}d\tau+
\int_{\tau_{n-1}}^{\ell}\mathbf{A}.\mathbf{n}d\tau \ . \label{JC:eq12}
\end{eqnarray}
For the first term of the right side, we can write
\begin{eqnarray}
\nonumber \int_{0}^{\tau_1}\mathbf{A}.\mathbf{n}d\tau =-H\int_{0}^{y_0/2}zdy  \ .
\end{eqnarray}
Since the equation for this segment of the trajectory is $z = D (1 - y(D \tan \theta
\sin \phi)^{-1}$, we get for the integral
\begin{eqnarray}
\int_{0}^{\tau_1}\mathbf{A}.\mathbf{n}d\tau=-\frac{HDy_0}{4} \ , y_0 = 2D \tan \theta
\sin \phi \ .
 \label{JC:eq13}
\end{eqnarray}
where $y_0$ is shown in Fig. \ref{mzbfig1}. Similarly, we find identical results for
the integrals of the vector potential over the other segments. Therefore for a
trajectory with the length $\ell$, the phase induced by the planar magnetic field
proportional to
\begin{eqnarray}
\Phi_{\ell}=\int_{0}^{\ell}\mathbf{A}.\mathbf{n}d\tau=\frac{-HDNy_0}{2} \ ,
 \label{JC:eq14}
\end{eqnarray}
where $N$ is the number of the triangles (each triangle consists of two segments) for
the trajectory of the length $\ell$ passing through the point $z$, $N =
[L/(4D\tan\theta\sin\phi)-z/(2D)] + [L/(4D\tan\theta\sin\phi)+z/(2D)]+2$, with the
square brackets denoting the integer part.

Thus, for the phase difference, we obtain
\begin{eqnarray}
\nonumber \varphi&=&\varphi_0+\pi\frac{\Phi}{\Phi_0}\frac{2N
D\tan\theta\sin\phi}{L}.\\
 \label{JC:eq15}
\end{eqnarray}
Here $\Phi=HDL$ is the total flux through the junction. Substituting this into Eq.
(\ref{JC:eq9}) and taking the $y$-component of the current, we obtain the final
expression
\begin{widetext}
\begin{eqnarray}
\nonumber \frac{I(\varphi_0)}{I_0}&=&
\frac{T}{T_c\xi_0}\sum_{\omega_{n}=-\infty}^{\omega_{n}=\infty}
\int_{0}^{D}\int_{0}^{\pi}\int_{-1}^{1}\frac{
\Delta_0^2(T)\sin\left(\varphi_0+\pi\frac{\Phi}{\Phi_0}
\frac{\ell}{L}(1-x^2)^{1/2}\sin\phi\right)(1-x^2)^{1/2}dx\sin\phi d\phi
dz}{(\Omega_{n}^2+\omega_{n}^2)\cosh \chi +2\Omega_{n}
\omega_{n}\sinh \chi+\Delta_{0}^{2}\cos\left(\varphi_0+
\pi\frac{\Phi}{\Phi_0}\frac{\ell}{L}(1-x^2)^{1/2}\sin\phi\right)},\\
I_c(\Phi)&=&max_{_{_{0\leq\varphi_0\leq2\pi}}}\frac{I(\varphi_0)}{I_0},
 \label{JC:eq16}
\end{eqnarray}
\end{widetext}
with the notations $\cos\theta=x$ and $I_0=2ev_{\rm F}\mathcal{N}(0)T_cW\xi_0$, and
$W$ being the width of the normal layer in the $x$-direction, which for a wide
junction is taken to be much larger than $D$ and $L$. The length $\ell$ of the
quasi-classical trajectory equals $\ell=2ND/x$. For low temperatures, $T \ll v_F/L$,
the summation over Matsubara frequencies in Eq. (\ref{JC:eq16}) can be replaced by
the integration,
$$2\pi T\sum(...)\rightarrow\int(...)d\omega\rightarrow\int(...)
\Delta_0(T)\cosh\mu d\mu \ ,$$ where $\omega=\Delta_0\sinh\mu$. For a long SNS
junction ($L\gg\xi_0$) and low temperature, Eq. (\ref{JC:eq16}) is simplified as
\begin{eqnarray}
\nonumber &&\frac{I(\varphi_0)}{I_0}=
\int_{0}^{D}\int_{0}^{\pi}\int_{-1}^{1}E(\varphi,\ell)(1-x^2)^{1/2}dx\sin\phi
d\phi dz,\\
 &&E(\varphi,\ell)=\frac{2D}{\xi_0}\left(\frac{\xi_0}{\ell}-
\frac{2\xi_0^2}{\ell^2}\right)\sum_{m=1}^{\infty}(-1)^m\frac{\sin(m\varphi)}{m},
 \label{JC:eq17}
\end{eqnarray}
where $E(\varphi,\ell)$ is the Fourier Series, given by (see Ref.
\onlinecite{GolubovRMP2004})
\begin{eqnarray}
 E(\varphi,\ell)=\frac{D}{\xi_0}\left(\frac{\xi_0}{\ell}-
\frac{2\xi_0^2}{\ell^2}\right) \bigg(
\varphi-2\pi\left[\frac{\varphi}{2\pi}+\frac{1}{2}\right] \bigg).
 \label{JC:eq18}
\end{eqnarray}

\section{Discussion and results
\label{sec3}}

Equation (\ref{JC:eq16}) expresses the magnetic interference
pattern $I_c(\Phi)$ of a ballistic SNS junction in the presence of
an in-plane magnetic flux $\Phi$. In this section we analyze
$I_c(\Phi)$ in terms of the length-to-thickness ratio $L/D$ and
the temperature $T$ for $D/\xi_0={\rm 1/30}$.
\begin{figure}
\centerline{\hbox{\epsfxsize=2.2in \epsffile {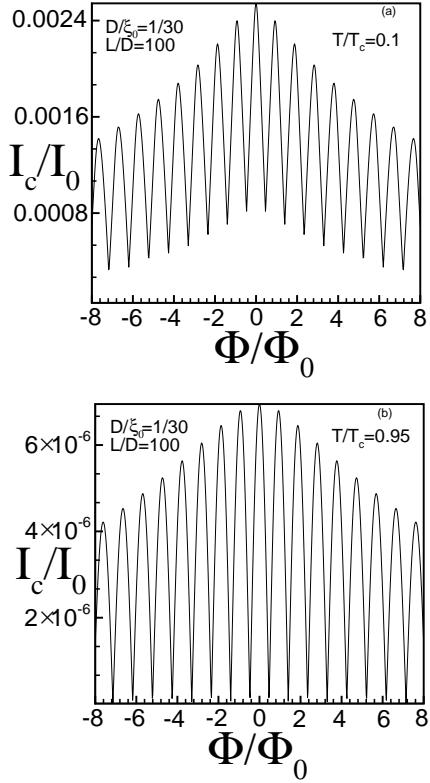}}} \caption{The critical
current dependence on the external magnetic field, applied in the plane of the normal
metallic layer for different temperatures, $L/D={\rm 100}$ and $D/\xi_0={\rm 1/30}$
.} \label{mzbfig2}
\end{figure}

Let us start with analyzing the case of very large $L/D$.  Figure
\ref{mzbfig2} a,b shows oscillations of $I_c(\Phi)$ for $L/D={\rm
100}$ and at low ($T=0.1T_c$) and high ($T=0.95T_c$) temperatures,
respectively. At low temperatures the the critical current goes
through nonzero minima at finite fluxes. The amplitude of
supercurrent minima decrease with $\Phi$ and drops to zero at
$\Phi\gg \Phi_0$. Compared to an ordinary magnetic interference
pattern, the oscillations are weakly damped since their amplitude
decreases with $\Phi$ much slower than as $~1/\Phi$. With
increasing temperature, the amplitude of the oscillations
decreases. Also, the minimal values of the supercurrent decreases
and vanishes as $T\rightarrow T_c$, where $\Delta_0(T) \ll T$.
Note that at both low and high temperatures the period of
oscillations varies from ${\rm 0.92}\Phi_0$ (first minima) at low
magnetic fluxes to $\Phi_0$ at high fluxes. The result that the
period of oscillations is temperature independent comes from the
fact that the gauge invariant phase in the argument of the sine
and cosine functions in Eq. (\ref{JC:eq16}) does not contain any
temperature-dependent factors.

\begin{figure}
\centerline{\hbox{\epsfxsize=2.2in \epsffile {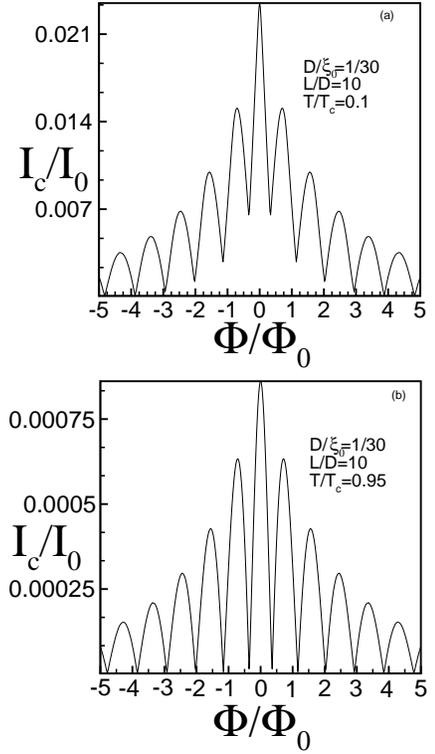}}}
\caption{The same as
Fig. \ref{mzbfig2} but for $L/D={\rm 10}$.} \label{mzbfig3}
\end{figure}

Figure \ref{mzbfig3} presents the magnetic interference pattern
for a lower $L/D={\rm 10}$ and at the same temperatures as Figure
\ref{mzbfig2}. From these plots we see that lowering $L/D$ has two
main effects. First, the rate at which the amplitude of $I_c$
oscillation decreases with $\Phi$ increases. Second, the period of
oscillations at both low and high temperatures becomes smaller
$\simeq {\rm 0.72}\Phi_0$ at small fluxes. Increasing $\Phi$ the
period increases up to $\simeq \Phi_0$. Again as in Figs.
\ref{mzbfig2} the value of supercurrent at the minima vanishes as
the temperature approaches $T_c$.

\begin{figure}
\centerline{\hbox{\epsfxsize=2.2in \epsffile {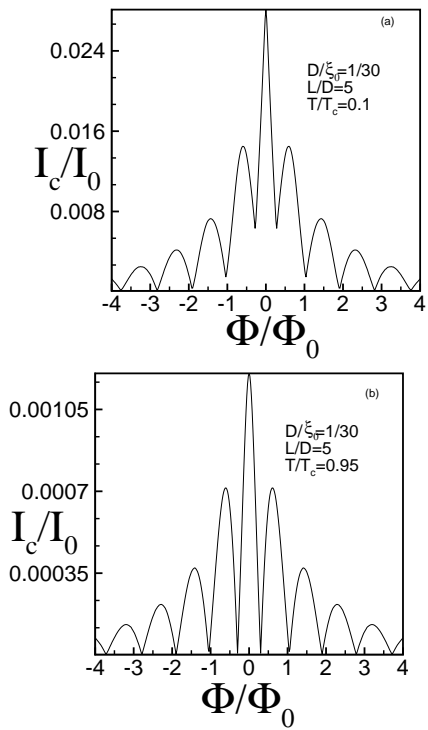}}}
\caption{The same as Fig. \ref{mzbfig2} but for $L/D={\rm 5}$.}
\label{mzbfig4}
\end{figure}

Still lower period of oscillations at small fluxes can be reached
at low values of $L/D$. This is illustrated in Fig. \ref{mzbfig4}
where the magnetic interference pattern is presented for $L/D={\rm
5}$. Clearly the decay is close to the ordinary pattern
(\ref{Fraunhofer}), {\em i.e.} $~1/\Phi$, and the period can be as
small as half the flux quantum.

The existence of the nonzero minima in the oscillations of $I_c(\Phi)$ is related to
the non-sinusoidal phase dependence of the Josephson current (\ref{JC:eq16}) which is
more pronounced at low temperatures $T\ll T_c$. We have found that $I_c(\Phi)$
undergoes a change of sign at a nonzero minimum. At such point the amplitude of the
first harmonic ($\propto \sin {\varphi_0}$) of the Josephson current vanishes and
$I_c(\Phi)$ is determined by the amplitude of the higher (mainly second) harmonics
which change sign upon crossing the minimum. Similar effect was found before in
ferromagnetic Josephson junctions (see Ref.
\onlinecite{SellierPRL2004}-\onlinecite{RyazanovPRB2006} and references therein). As
$T$ approaches $T_c$ the ratio $\Delta_0(T)/T$ goes to zero and the current-phase
relation (Eq. (\ref{JC:eq16})) becomes sinusoidal and consequently the nonzero minima
disappear.

The dependence of the period of the oscillations on the magnetic flux and the
geometry can be understood in terms of the difference between  the magnetic flux
$\Phi_{\ell}$ (Eq. (\ref{JC:eq14})) enclosed by a trajectory of length $\ell$, and
half of the flux $\Phi$ penetrating through the area $DL$. The difference comes from
the fact that a trajectory which does not pass through the edges of S contacts has
extra parts in N-layer which lie outside the area $DL$ (See Fig. \ref{mzbfig1}).
Writing $\Phi_{\ell}=\Phi /2+\delta \Phi_{\ell}$, the difference $\delta \Phi_{\ell}$
vanishes only for the trajectories which pass through the edges of S contacts. A
finite averaged $\langle \delta \Phi_{\ell} \rangle$ over different
trajectories means that the
period of $I_c(\Phi)$-oscillations, obtained from Eq. (\ref{JC:eq16}), differs from
$\Phi_0$. We note that in the limit of thin N-layer $L \gg D$ or high magnetic fluxes
$\Phi \gg \Phi_0$ the contribution of the trajectories which are not passing through
the edges is negligibly small in the interference structure and the period of the
oscillation approaches $\Phi_0$ (See Figs. \ref{mzbfig2}-\ref{mzbfig4}).

In contrast to $L\gg D$ case, for smaller $L/D$ the contribution
of the trajectories not passing through the edges is important.
Because of having larger length, a trajectory not passing through
the edges has bigger contribution in gaining the effect of the
magnetic flux as compared to the corresponding trajectory (having
the same orientation $\theta$ and $\phi$) passing through the
edges. Therefore we expect that the effect of magnetic flux is
more pronounced for thicker N layers compare to the thinner ones,
which explains why the decrease of the amplitude of
$I_c(\Phi)$-oscillations with $\Phi$ is faster for smaller $L/D$.

\section{Conclusions \label{sec4}}

In conclusion, we have studied Josephson effect in a SNS structure
made of a thin ballistic N-layer of thickness $D$ on which two
superconducting electrodes are deposited at the distance $L$  between
each other.  A magnetic field is applied in plane of the N-layer
which modulates the superconducting interference and leads to a
decaying oscillatory variation of the critical supercurrent
$I_c(\Phi)$ with the magnetic flux $\Phi$. Using the
quasiclassical Green's functions approach, we have shown that such
a magnetic interference pattern has three main differences with
that of an ordinary pattern. First, at low temperatures the
oscillations of the critical current $I_c(\Phi)$ go through the minima
at which the supercurrent has nonzero values. Second, for a large $L/D$ the
amplitude of the quasi-periodic oscillations of $I_c(\Phi)$ decays
at a rate which is much slower than $1/\Phi$. Third, at low
magnetic fluxes the oscillations can have a period smaller than
the magnetic quantum flux $\Phi_0$ depending on $L/D$. These features
have been experimentally observed recently~\cite{KlapwijkNature2006}.

\begin{acknowledgments}
We acknowledge useful discussions with  D. Huertas-Hernando, A. G.
Moghaddam and Yu. V. Nazarov. M. Z. thanks G. E. W. Bauer for the
hospitality and support during his visit to Kavli Institute of
NanoScience at Delft where this work was initiated. This work was
supported in part by EC Grant No. NMP2-CT2003-505587 (SFINX).
\end{acknowledgments}

\end{document}